\pdfoutput=1

\documentclass{acm_proc_article-sp}

\graphicspath{{../figures/}{../data/}}
\DeclareGraphicsExtensions{.pdf,.png,.jpg,.eps}

\usepackage{algorithm, algorithmic, array, subcaption, fixltx2e, url}
\def\sharedaffiliation{%
\end{tabular}
\begin{tabular}{c}}
\begin{document}

\title{A Simple Multipath OpenFlow Controller using topology-based algorithm for Multipath TCP}
%
%
%
%
%

\numberofauthors{4} 
%
\author{
%
%
    \alignauthor Chawanat Nakasan\\
    \affaddr{Nara Institute of Science and Technology}\\
    \affaddr{8916-5 Takayama, Ikoma}\\
    \affaddr{Nara 630-0101, Japan}\\
    \email{chawanat.nakasan.cb5@
is.naist.jp} \\
    \alignauthor Kohei Ichikawa \\
    \affaddr{Nara Institute of Science and Technology}\\
    \affaddr{8916-5 Takayama, Ikoma}\\
    \affaddr{Nara 630-0101, Japan}\\
    \email{ichikawa@is.naist.jp} \\
    \alignauthor Hajimu Iida \\
    \affaddr{Nara Institute of Science and Technology}\\
    \affaddr{8916-5 Takayama, Ikoma}\\
    \affaddr{Nara 630-0101, Japan}\\
    \email{iida@itc.naist.jp} \\
    \sharedaffiliation
    \and
    \alignauthor Putchong Uthayopas\\
    \affaddr{Kasetsart University}\\
    \affaddr{50 Phahonyothin Rd., Lat Yao, Chatuchak}\\
    \affaddr{Bangkok 10900, Thailand}\\
    \email{pu@ku.ac.th}\\
}

\maketitle
\begin{abstract}
Multipath TCP, or MPTCP, is a widely-researched mechanism that allows a single application-level connection to be split to more than one TCP stream, and consequently more than one network interface, as opposed to the traditional TCP/IP model. Being a transport layer protocol, MPTCP can easily interact between the application using it and the network supporting it. However, MPTCP does not have control of its own route. Default IP routing behavior generally takes all traffic through the shortest or best-metric path. However, this behavior may actually cause paths to collide with each other, creating contention for bandwidth in a number of edges. This can result in a bottleneck which limits the throughput of the network. Therefore, a multipath routing mechanism is necessary to ensure smooth operation of MPTCP. We created smoc, a Simple Multipath OpenFlow Controller, that uses only topology information of the network to avoid collision where possible. Evaluation of smoc in a virtual local-area and a physical wide-area SDNs showed favorable results as smoc provided better performance than simple or spanning-tree routing mechanisms.
\end{abstract}




\section{Introduction}
Networked systems, such as distributed database, computation, and file storage, have become more complex with increasing capacity. This evolution increases demands on the network. Many practices have been developed to improve their functionality or alleviate problems, such as \textit{multi-homing} which connects a system to the Internet through multiple gateways. Another concept, \textit{multi-site}, refers to the practice of distributing the system to multiple geographic locations. These practices have many benefits including locality, capacity, and redundancy. When these two concepts are used together, the sites of the networked system can be connected through wide-area network (WAN) by multiple paths.

However, having multiple paths between the sites does not mean both paths are always used. In a traditional network model, one application layer socket is supported by one transport-layer session \cite{dong07}, which is supported by a fixed pair of network-layer and link-layer endpoints \cite{rfc1180}. \textit{Multipathing} was developed to allow a host or network to utilize multiple paths at the same time. Multipathing has many benefits such as increasing maximum available bandwidth, balancing network load, and providing redundancy. Multipathing with commodity network medium can also be used as a low-cost alternative to using a single expensive network medium. Many multipathing solutions exist, with their own set of benefits and problems.

\textit{Multipath TCP} (MPTCP) \cite{rfc6182, rfc6824} is a promising multipathing protocol which was developed as an extension to TCP. As an extension, MPTCP was designed to be backwards-compatible with current applications and networks. Additionally, being a transport layer protocol based on TCP, it has congestion control, making it useful in WANs which can be (relatively to LANs) unequal and unstable.

Even though MPTCP has many advantages, it also has a major shortcoming. As a transport layer protocol, it has no control over its own routing. Routing problems can arise if sites on a network are connected through a shared infrastructure and the routing system works in a legacy manner. Without the knowledge that MPTCP is being used, the network may route the multiple MPTCP ``subflows" through the same path causing a bottleneck, while also leaving some other paths unused causing underutilization. When this is the case, the benefits of MPTCP would be limited due to the bottleneck and underutilization. Spreading MPTCP traffic across multiple paths become an important topic because its usefulness could be improved.

The role of a customizable routing mechanism that would be suitable for developing a routing mechanism for MPTCP would be easily filled in by OpenFlow \cite{openflow}, a software-defined network (SDN) protocol.

In this work, we aim to create a simple and efficient OpenFlow controller that splits and distributes MPTCP traffic through the network. This would increase bandwidth utilization of MPTCP in the network so that full capacity may be used. We have three core ideas behind our controller design. Firstly, the controller should be backwards-compatible with non-MPTCP traffic. Secondly, we strive to increase bandwidth available to an application. Finally, we attempt to explore various multipath routing strategies; one simple strategy is presented in this paper.

Our work is primarily targeted at large-scale, multi-homed multi-site systems connected together using OpenFlow. This class of systems include distributed storage, database, content delivery network (CDN), high-performance computing (HPC), or systems with disaster recovery (DR) sites, which are usually far away from the main site and connected through WANs. Since bandwidth in WAN is more limited compared to local-area networks, improved efficiency has greater impact in this kind of network.

\begin{figure*}[!t]
    \centering
    \begin{subfigure}[t]{0.25\textwidth}
        \centering
        \includegraphics[height = 3.5 cm]{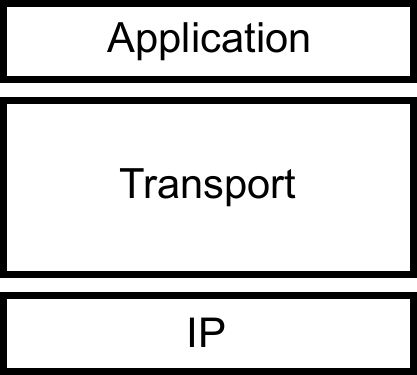}
        \caption{}
        \label{fig:layers_tcpip}    
    \end{subfigure}\hfil
    \begin{subfigure}[t]{0.25\textwidth}
        \centering
        \includegraphics[height = 3.5 cm]{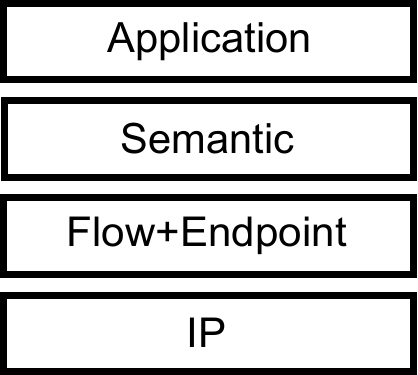}
        \caption{}
        \label{fig:layers_logjam}    
    \end{subfigure}\hfil
    \begin{subfigure}[t]{0.45\textwidth}
    	\centering
        \includegraphics[height = 3.5 cm]{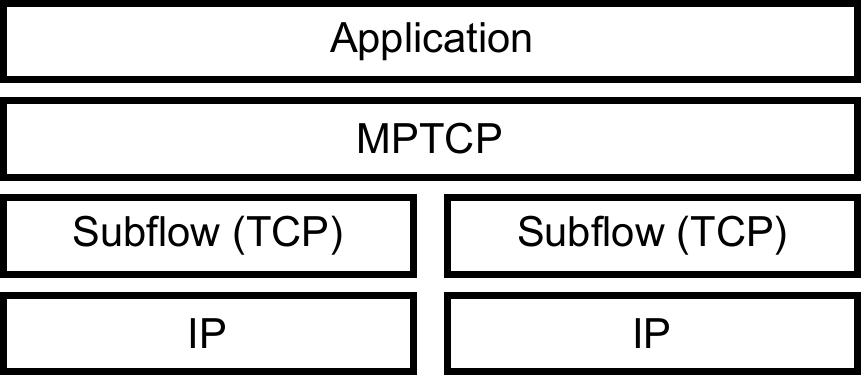}
        \caption{}
        \label{fig:layers_mptcp}
    \end{subfigure}
    \caption{Relationship between the traditional TCP/IP network model and derivation that led to the principle design of MPTCP. (a) shows the regular TCP/IP model, (b) is the decomposition of transport layer which is the basis of MPTCP, and (c) is the actual MPTCP implementation of this model.}
\end{figure*}

\section{Background}

\subsection{Software-defined Network}
\textit{OpenFlow} is an SDN protocol which allows network traffic control and management from the centralized OpenFlow Controller. Centralization allows the  network elements to be programmed to add or remove any switching or routing rules in its \textit{flow table}. While OpenFlow provides programming flexibility to the network and allows many concepts such as QoS or traffic engineering to be realized, it cannot modify communication pattern between the end hosts. In traditional TCP/IP protocol suite, only one route or path is used per connection. This limits the maximum bandwidth to only one path, and not that of the entire network. This limitation cannot be circumvented by OpenFlow.

\subsection{Multipathing}
Multipathing allows us to use more than one path in one logical connection, increasing bandwidth utilization, improving redundancy and stability, as well as allowing seamless handovers in certain environments. It is especially useful in multi-homed systems, which see a recent upward trend. Multipathing has been attempted from many perspectives for various purposes. We can roughly classify them into three general classes based on TCP/IP model layer as follows.

\subsubsection{Application Layer}
In the application layer, multipathing can be done by creating multiple sockets from the application. One notable example, GridFTP \cite{allcock2000gridftp, allcock2005globus}, uses multiple TCP streams in parallel to improve performance along the network topology \cite{gunter2012parallelism} by extending FTP to support parallel streams. However, application layer multipathing can be error-prone \cite{barre10} and hard to maintain \cite{dong07}. This is because the task of working with the multiple paths and flows will fall upon the application, which is not aware of the many mechanisms that are already available and working in the transport layer \cite{jacobson88}. For example, the application may not be aware of unequal paths and continue to push equal data to both sockets, causing traffic to stay behind in the slower path.

\subsubsection{Network Layer}
Equal-Cost Multipath (ECMP) \cite{rfc2991, rfc2992} is a multipathing mechanism that allows multiple paths to be used on the network layer when they have equal costs. While it is simple and efficient as it can quickly select paths based on the packet header, TCP is not aware of ECMP. Some ECMP path selection and hashing strategies (such as simple round-robin) may cause packets to arrive out-of-order or unbalance the network, prompting TCP to retransmit as multiple duplicate ACKs may be received, resulting in decreased network performance \cite{dong07}. Additionally, ECMP is designed for \textit{equal}-cost networks and therefore will not work when path costs are not equal, such as in WANs, due to variation in bandwidth and latency.

\subsubsection{Transport Layer}
Transport layer is more informed about each path's conditions than the applications \cite{barre10} and also more aware of high-level connections than the network layer. One prominent example is Stream Control Transmission Protocol (SCTP) \cite{rfc4960}. However, while SCTP is also capable of multipathing, the feature is aimed for redundancy, not bandwidth utilization. Additionally, middleboxes such as network address translators (NATs) are not aware of SCTP and may block it. Applications also need to explicitly use SCTP because it is a completely new protocol, presenting a further compatibility problem \cite{rfc6182}.

To address this compatibility problem, protocols such as \textit{concurrent TCP (cTCP)} \cite{dong07} and \textit{M/TCP} \cite{kultida04} (along with others mentioned in \cite{chihani11}) are based on or compatible with TCP. Among these protocols, MPTCP is one of the most promising as it has rich features, backwards-compatibility with TCP, and extensive research, including a Linux kernel implementation \cite{raiciu12}.

\subsection{MPTCP}

\textit{Multipath Transmission Control Protocol,} or \textit{MPTCP} is an extension to \textit{TCP} at the transport layer that utilizes multiple paths between two network endpoints by stripping data into multiple \textit{subflows}. Each subflow behaves like a TCP flow, with its own congestion control, send and receive windows, and so on.
MPTCP interface for applications is a complete drop-in replacement, meaning that the applications need not be modified. An MPTCP session would be created for each socket opened by the application. For example, opening 5 sockets to download a file from an HTTP server would result in 5 separate MPTCP sessions being opened for the download application.

By decomposing the transport layer into two sublayers, as shown from Figure \ref{fig:layers_tcpip} to Figure \ref{fig:layers_logjam} \cite{ford08}, MPTCP can separately recognize end-to-end and point-to-point situations better than the traditional model by having the upper half, which is the MPTCP extension, manage the connections and subflows, while the lower half works with congestion and other matters in each subflow as in Figure \ref{fig:layers_mptcp}.

\subsection{Our interpretation of {\subsecit multipathing}}
Although reviewed literature does not give a clear definition for the term \textit{multipathing}, they generally agree that it means \textit{creating multiple network connections or sessions between a pair of hosts, with the connections or sessions traveling through different paths (when available) across the network}. We will use this meaning in our work.

\section{Design of MPTCP Routing Mechanism}
Two actions are necessary to route MPTCP traffic through the network using multiple paths. First, we need to know which subflows belong to which instance of MPTCP. Second, we also need to decide which paths would be used and when. These actions are further discussed in the following two subsections.

\subsection{Identifying MPTCP subflow group on the network layer}
\begin{figure}[!t]
    \centering
    \includegraphics[width=0.9\linewidth]{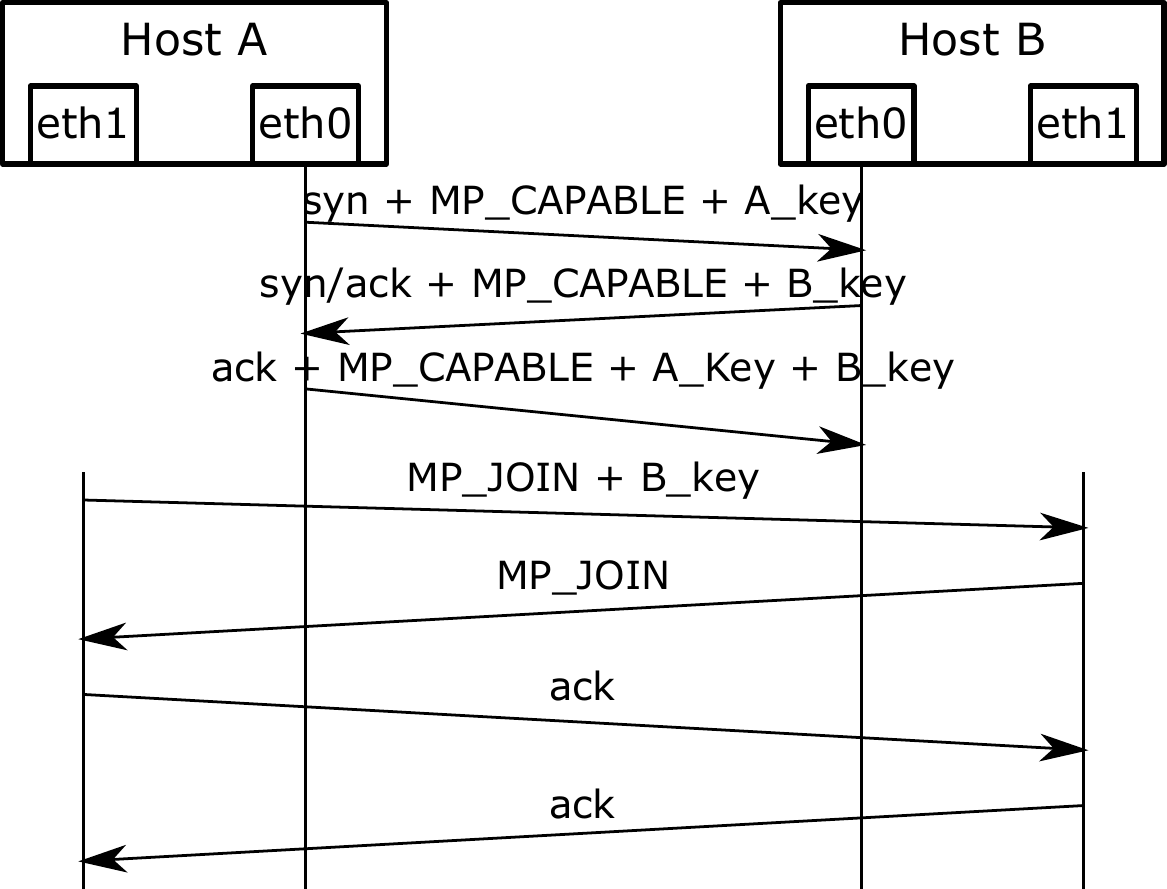}
    \caption{MPTCP handshake process, annotated with a partial list of TCP and MPTCP option fields used in our work.}
    \label{fig:mptcp_handshake}
\end{figure}

As stated above, we need to identify which subflows belong to the same MPTCP instance. Since all MPTCP information is encoded as TCP options, not headers, it is impossible to simply use OpenFlow's normal matching methods to identify MPTCP subflow grouping. Therefore, a method to identify the subflows from the OpenFlow controller's perspective is necessary. In order to do so, special information beyond IP addresses and TCP port numbers are required.

Fortunately, MPTCP exchanges all needed information during the initial MPTCP subflow establishment (using \texttt{MP\_CAPA{\allowbreak}BLE} TCP option) and subsequent subflows (using \texttt{MP\_JOIN} option). MPTCP relies on \textit{keys} and \textit{tokens} to identify a connection endpoint which is unique for each connection and host. We can use this identification information to find which subflows belong to which MPTCP connection\footnote{In MPTCP, keys are later hashed, truncated, and called \textit{tokens}. As we do not work on the full process of MPTCP, we will not care about the differences between these terms. \textit{Key} will be used throughout this paper for simplicity.}. When MPTCP creates a new instance for the first time, each host sends its own key to the other host. When a host establishes an additional subflow, it (A) will send the other party (B)'s key to identify an MPTCP session it (A) wishes to join. As this process uses different IP address and TCP port pairs, an OpenFlow \texttt{packet\_in} message will be sent from an OpenFlow switch to the controller, which would use this information. This process is illustrated in Figure~\ref{fig:mptcp_handshake} and is used as a basis for flow detection and grouping in our routing algorithm.


\subsection{Finding and using multiple paths}
Apart from correctly identifying MPTCP subflows, we need to know which paths each flow should take. We aim to create a suite of MPTCP and OpenFlow working cooperatively in the same system implementation. In this work, OpenFlow would find optimal \textit{path sets}, a collection of paths that lead a packet from one host to another, and decide which path an MPTCP subflow should use. This mechanism involves multiple stages: first we analyze the packet and gather or match information with the database, we may need to find a new path set if necessary, then the path set must be applied to new MPTCP subflows as they are created. By cycling through the different paths in a path set, MPTCP subflows can be distributed to multiple paths.

\subsubsection{Path Set Calculation Algorithm}
\label{subsec:algorithms}

\begin{algorithm}
\caption{Algorithm to find a path set to route from S1 to S2 in network graph G}
\label{alg:pathset}
\begin{algorithmic}
\REQUIRE graph $G(V,E)$
\REQUIRE $S1$, $S2$ $\in$ $switches$
\STATE $G(V,E)$ $\leftarrow$ $NetworkTopology(switches, links)$
\STATE $PrimaryPath$ $\leftarrow$ shortest\_path($G$, $S1$, $S2$)
\STATE $AltPaths$ $\leftarrow$ all\_simple\_paths($G$, $S1$, $S2$) - $PrimaryPath$
\STATE $AltPaths$ $\leftarrow$ $AltPaths$ sorted by number of edges shared with PrimaryPath ascending, by length of path ascending
\STATE $PathSet$ $\leftarrow$ $PrimaryPath + AltPaths$
\RETURN $PathSet$
\end{algorithmic}
\end{algorithm}

\begin{algorithm*}[tb]
\caption{Algorithm to handle incoming MPTCP packets that trigger OpenFlow packet-in message}
\label{alg:tables}
\begin{algorithmic}
\REQUIRE $packet$
\ENSURE $route$ to route the flow $packet$ belongs to
\STATE $pending\_capable$ $\leftarrow$ hash((init\_ip\_port, listen\_ip\_port) $\rightarrow$ (init\_key, pathset))
\STATE $pending\_join$ $\leftarrow$ hash((init\_ip\_port, listen\_ip\_port) $\rightarrow$ (listen\_key, pathset))
\STATE $mptcp\_connections$ $\leftarrow$ hash(dst\_key $\rightarrow$ (src\_key, pathset))
\IF{$packet$ is \texttt{MP\_CAPABLE} message}
  \IF{($packet.dst\_ip\_port$, $packet.src\_ip\_port$) in $pending\_capable$}
    \STATE $recvkey$, $ABpathset$ $\leftarrow$ $pending\_capable$[($packet.src\_ip\_port$, $packet.dst\_ip\_port$)]
    \STATE $BApathset$ $\leftarrow$ find new pathset
    \STATE add $recvkey$ $\rightarrow$ ($packet.sendkey$, $ABpathset$) to $mptcp\_connections$
    \STATE add $packet.sendkey$ $\rightarrow$ ($recvkey$, $BApathset$) to $mptcp\_connections$
    \STATE delete key ($packet.dst\_ip\_port$, $packet.src\_ip\_port$) from $pending\_capable$
    \RETURN $BApathset$.next()
  \ELSE
    \STATE $ABpathset$ $\leftarrow$ find new pathset
    \STATE add ($packet.src\_ip\_port$, $packet.dst\_ip\_port$) $\rightarrow$ ($packet$.$sendkey$, $ABpathset$) to $pending\_capable$
    \RETURN $ABpathset$.next()
  \ENDIF
\ELSIF{$packet$ is \texttt{MP\_JOIN} message}
  \IF{($packet.dst\_ip\_port$, $packet.src\_ip\_port$) in $pending\_join$}
    \STATE $sendkey$, $ABpathset$ $\leftarrow$ $pending\_join$[($packet.src\_ip\_port$, $packet.dst\_ip\_port$)]
    \STATE $recvkey$, $BApathset$ $\leftarrow$ $mptcp\_connections$[$sendkey$]
    \STATE delete key ($packet.dst\_ip\_port$, $packet.src\_ip\_port$) from $pending\_join$
    \RETURN $BApathset$.next()
  \ELSE
    \STATE $sendkey$, $ABpathset$ $\leftarrow$ $mptcp\_connections$[$packet$.$recvkey$]
    \STATE add ($packet.src\_ip\_port$, $packet.dst\_ip\_port$) $\rightarrow$ ($sendkey$, $ABpathset$) to $pending\_join$
    \RETURN $ABpathset$.next()
  \ENDIF
\ENDIF
\RETURN shortest path as $route$ (otherwise)
\end{algorithmic}
\end{algorithm*}

Algorithm~\ref{alg:pathset} describes a simple method to find a path set for multipath use. When supplied with a network topology graph and the source and destination switches, the algorithm chooses one shortest path as the \textit{primary path}. The remaining paths are sorted and prioritized to minimize path sharing with the primary path, and then by path length. We used the shortest path and all simple path functions from \texttt{networkx}\cite{hagberg2008exploring} package. The path sets are stored as Python \texttt{itertools.cycle} object, which allows us to easily cycle through all paths inside.

\subsubsection{Collecting and Managing MPTCP Subflows}
To maintain the states of MPTCP subflows, we use three tables to store and match the subflows and assign them to routes by Algorithm \ref{alg:tables}:\begin{enumerate}
\item \texttt{pending\_capable} table stores information of first SYN packets sent by the MPTCP initiator using \texttt{MP\_CAPABLE} message. It maps the IP address and TCP port to the initiator's hash and also stores the path set from the initiator to the listener.
\item \texttt{pending\_join} table does a similar function for subsequent subflows created by \texttt{MP\_JOIN} messages.
\item \texttt{mptcp\_connections} table stores established MPTCP connections. Once an entry in the previous two tables is matched by a reply packet (TCP ACK), that entry is removed from its original table and the path set will be stored here. It maps a destination's key to the source's key and the path set from source to destination.
\end{enumerate}

\section{Implementation of smoc: Simple Multipath OpenFlow Controller}
\begin{figure}
    \centering
    \includegraphics[width=\linewidth]{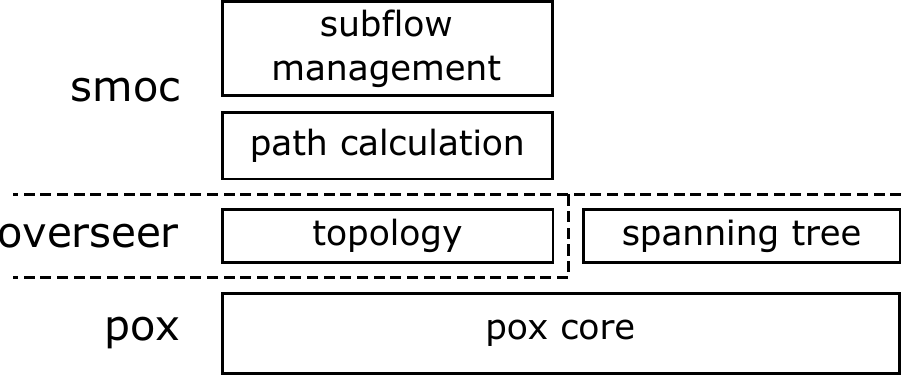}
    \caption{Components of the \texttt{smoc} controller, based on POX framework and Overseer's topology management modules.}
    \label{fig:smoc}
\end{figure}

To achieve our goal of solving the multipath bottleneck problem by using OpenFlow to route MPTCP, we implemented the algorithms described in Subsection~\ref{subsec:algorithms} in our controller, Simple Multipath OpenFlow Controller (smoc). The core of smoc is based on POX, a well-known OpenFlow framework. POX was chosen due to its modularity which means new features can be rapidly developed. Topology management and path management features are based on Overseer \cite{uchupala2014sdn, uchupala2014application} which is also an OpenFlow controller based on POX. Overseer's original purpose is to optimize routing based on characteristics of applications. To serve its purpose, Overseer has well-designed topology management and path management features, which also form the basis for smoc.

Path finding is assisted by the \texttt{networkx} Python package while path selection is based on Algorithm~\ref{alg:pathset}. smoc uses Algorithm~\ref{alg:tables} to manage all subflows and incoming packets. Underlying maintenance functions such as spanning tree management and OpenFlow protocol are handled by POX and Overseer.

To route flows, we maintain a list of pending and connected sessions. When we receive a new connection handshake message, we calculate a new path set and add both the information of the connection initiation and the path set to the pending list. When the pending connection is responded, we calculate another path set for the reverse direction and move everything to the connected sessions list. Any subsequent connections would only require a lookup in the connected sessions list to find an appropriate path.

\section{Evaluation and Results}
\begin{figure}
    \centering
    \begin{subfigure}{\linewidth}
        \centering
    \includegraphics[width=\linewidth]{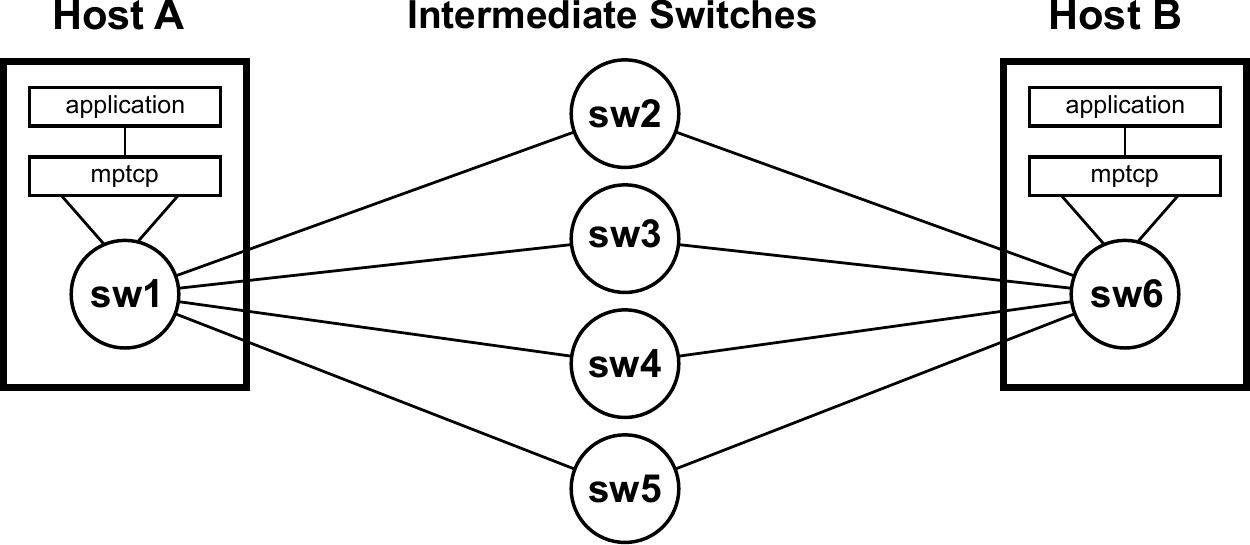}
    \caption{Topology 1 on local testbed}
    \label{fig:topo_i}
    \end{subfigure}\hfil
    \begin{subfigure}{\linewidth}
        \centering
    \includegraphics[width=\linewidth]{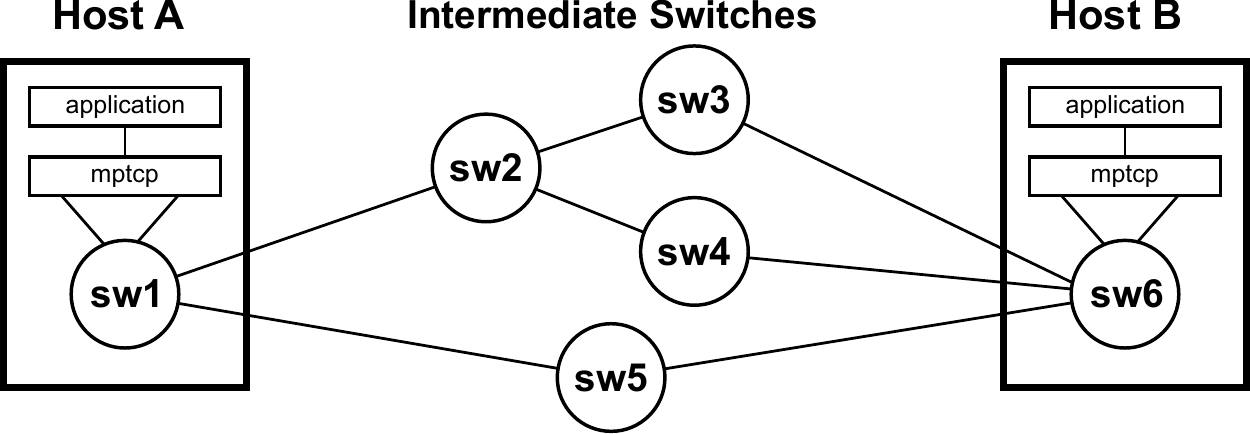}
    \caption{Topology 2 on local testbed}
    \label{fig:topo_j} 
    \end{subfigure}\hfil
    \caption{Topology configuration of the local testbed. The switches are Open vSwitch installed on virtual machines. Each virtual machine is hosted on a separate physical host. Links between the switches are limited to 100 Mbps.}
    \label{fig:topo_local}
\end{figure}

\begin{figure}[tb]
    \includegraphics[width=\linewidth]{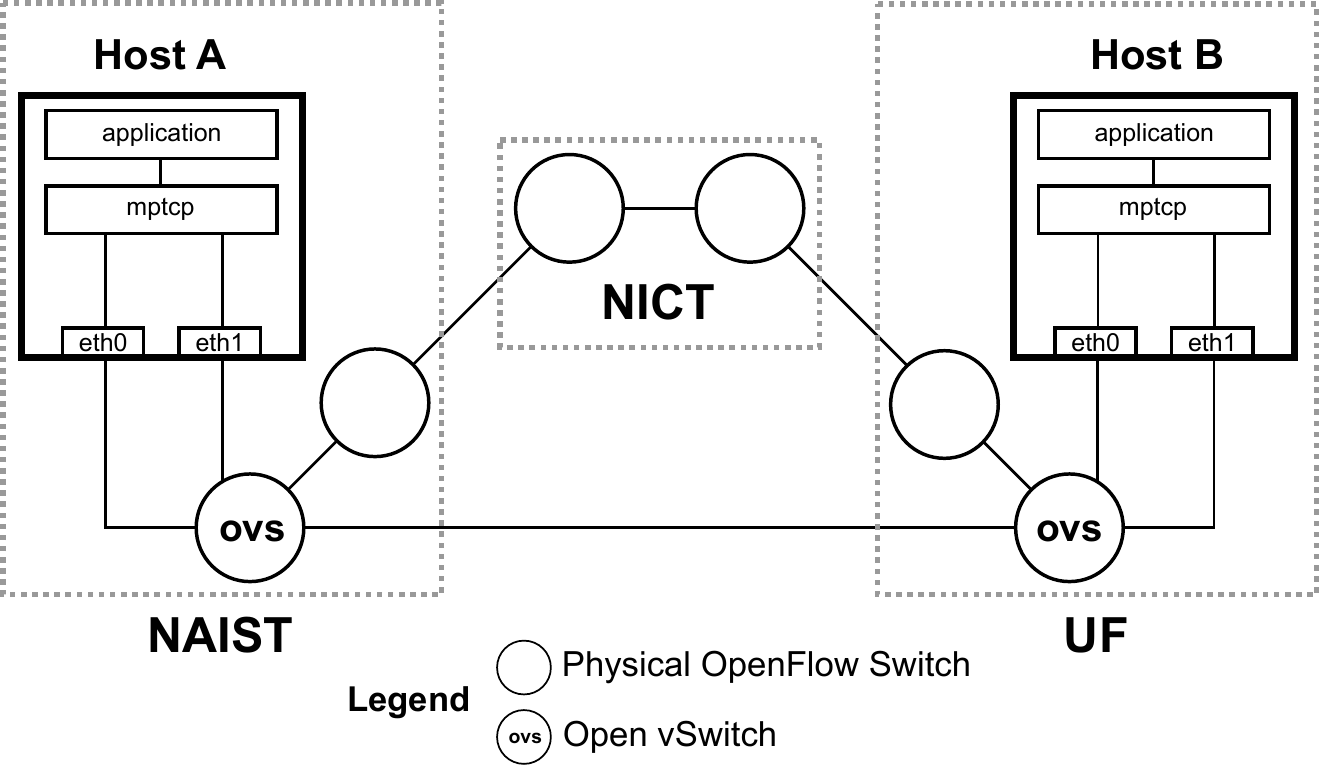}
    \caption{Testbed implementation in PRAGMA-ENT. The hosts are installed as virtual machines on the two sites.}
    \label{fig:testbed_pragma}
\end{figure}

\begin{figure}
    \centering
    \begin{subfigure}{\linewidth}
        \centering
    \includegraphics[width=\linewidth]{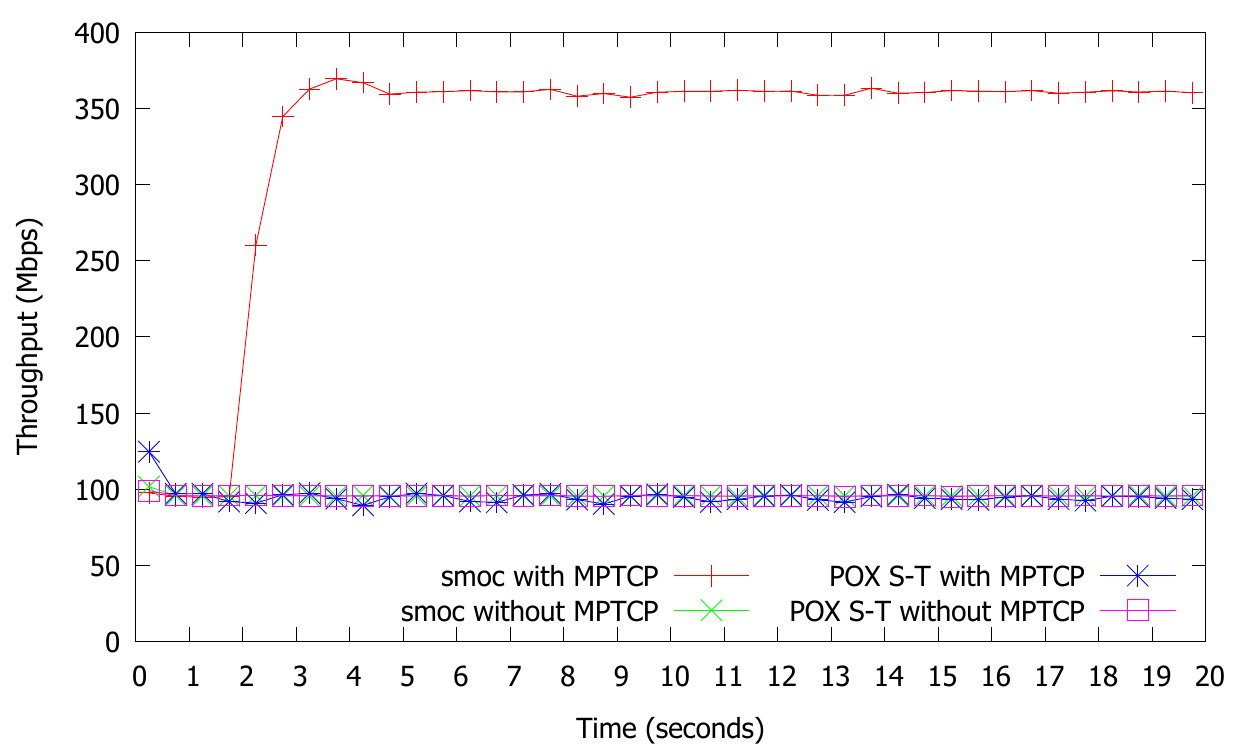}
    \caption{Results of test on local testbed with Topology 1}
    \label{fig:plot_isa1}
    \end{subfigure}\hfil

    \begin{subfigure}{\linewidth}
        \centering
    \includegraphics[width=\linewidth]{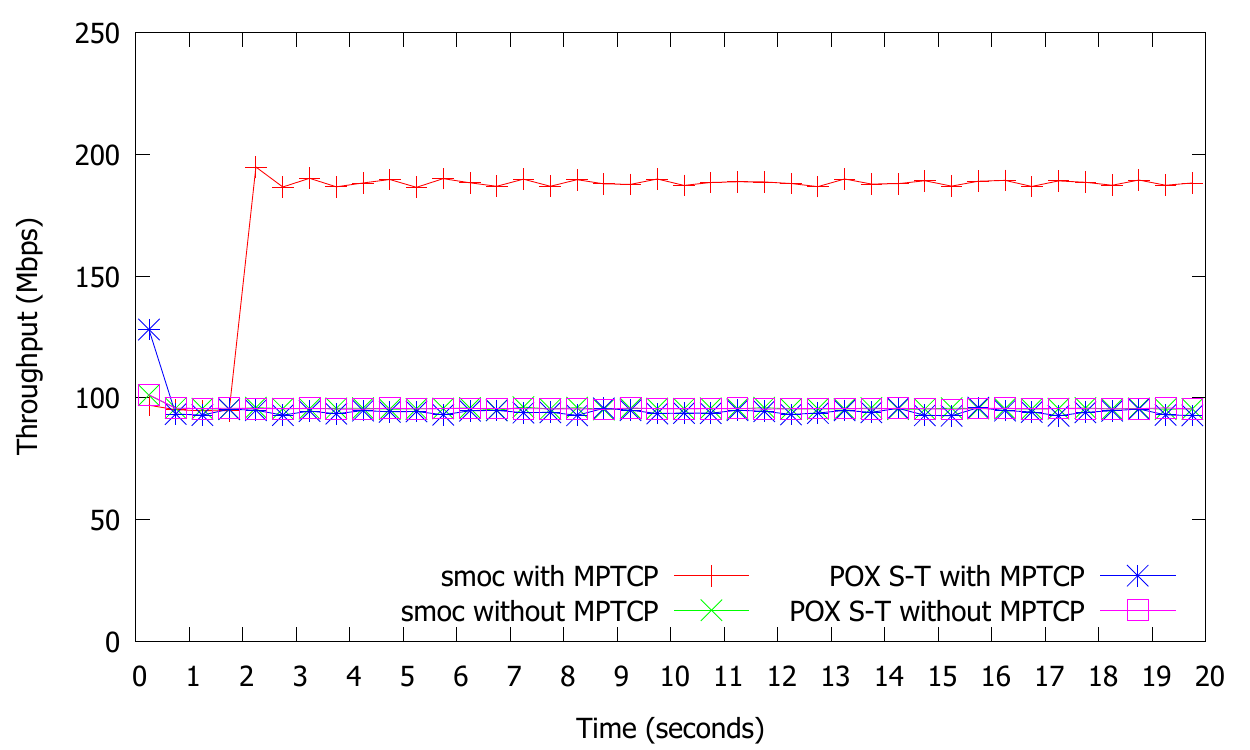}
    \caption{Results of test on local testbed with Topology 2}
    \label{fig:plot_isa2} 
    \end{subfigure}\hfil

    \begin{subfigure}{\linewidth}
        \centering
    \includegraphics[width=\linewidth]{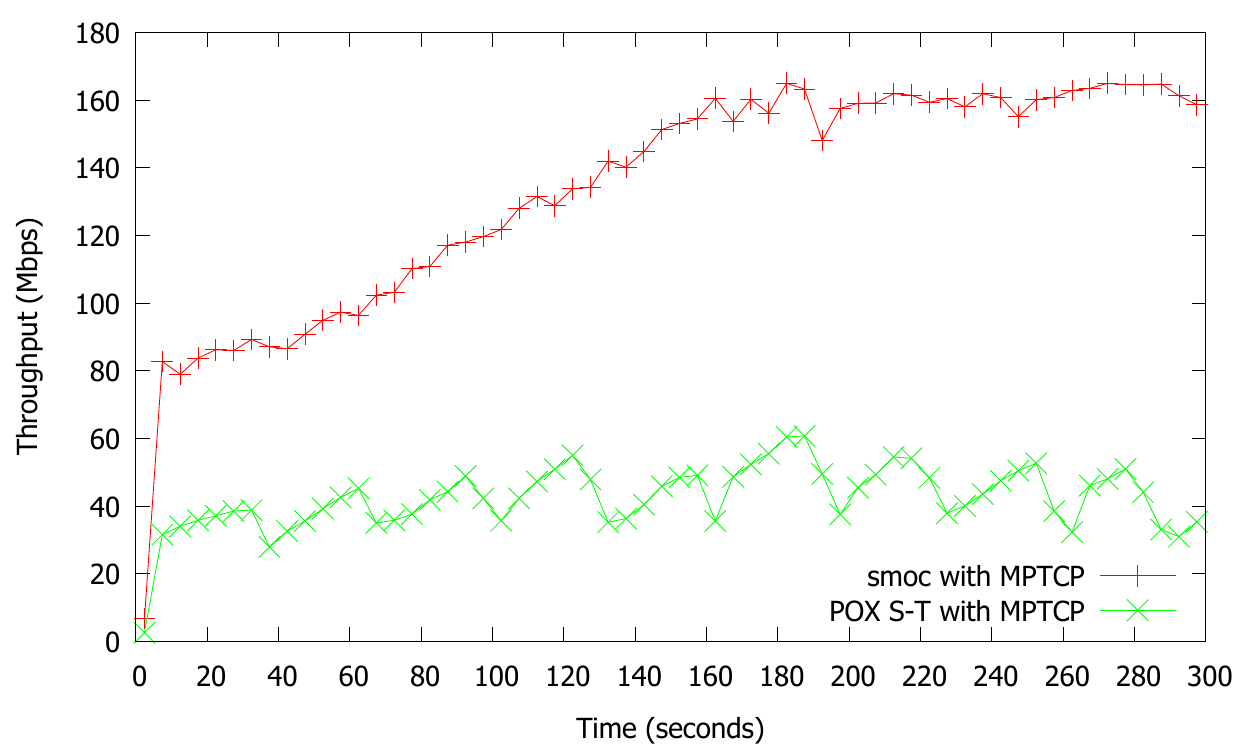}
    \caption{Results of test on wide-area testbed with PRAGMA-ENT}
    \label{fig:plot_ent} 
    \end{subfigure}\hfil

    \caption{Transient throughput between two hosts measured by \texttt{iperf} on different network topologies}
    \label{fig:plot_isa}
\end{figure}
We evaluated \texttt{smoc} against POX's original spanning tree controller (henceforth, POX S-T).
With this controller, all MPTCP traffic would be confined to a single path even if multiple paths actually exist in the network.
We chose this controller because it is based on the same framework and architecture, and spanning tree is commonly used to prevent loops in network topology. However, spanning tree eliminates any sort of multiple paths that exist at the network topology level. This means POX S-T always produces a single path between any pair of hosts. Being based on the same technology as smoc, all basic program libraries would be the same. This makes POX S-T suitable for an experimental control.

We chose \texttt{iperf} as our benchmarking tool due to its simplicity. smoc was evaluated in two testbeds, a local- and a wide-area testbed, which represented different network environments.

In the local testbed, two topology configurations shown in (Figure~\ref{fig:topo_local}) are modeled after a previous work from our research group \cite{huang2015multipath}. Topology 1 (Figure~\ref{fig:topo_i}) has four isolated paths, while Topology 2 (Figure~\ref{fig:topo_j}) has paths partly sharing a link.

The wide-area testbed (Figure~\ref{fig:testbed_pragma}) experiment uses an existing collaborative wide-area software-defined network project known as the Pacific Rim Applications and Grid Middleware Assembly Experimental Network Testbed (PRAGMA-ENT)~\cite{pragmaexp}.

\subsection{Evaluation in virtual local-area SDN}~
We implemented our local-area testbed on a VMware vSphere environment using six virtual machines. Each virtual machine, containing MPTCP installation and Open vSwitch \cite{ovs}, were deployed to different physical host machines. The GRE connections established between each virtual machine are manually limited to 100 Mbps to ensure that our virtual environment has a stable and clear maximum level of bandwidth, allowing easier verification of MPTCP and our controller.

We obtained MPTCP kernel and utilities from \cite{mptcpkernel}. The kernel provides MPTCP functionality while the other MPTCP utilities allow us to disable MPTCP on select interfaces to make sure that the experimental traffic does not ``spill'' into the management subnet.
This MPTCP kernel comes with multiple options that can be set through the \texttt{sysctl} variables, allowing us to customize the subflow creation options and numbers. Some options allow an arbitrary number of subflows to be created, regardless of the actual number of interfaces of the machine.

Testing POX S-T and smoc produced results as shown in Figure~\ref{fig:plot_isa}. Without a combination of a multipath router and MPTCP, only one path could be used at a time and the test run showed only approximately 95 Mbps of bandwidth, slightly below the theoretical limit was 100 Mbps, was used. However, when smoc and MPTCP are used together, after a few seconds of delay in the controller, the measured bandwidth was increased to greater than 100 Mbps, indicating that multipathing was successful with this combination. Test results using Topology 1, shown in Figure~\ref{fig:plot_isa1}, indicate that all four paths between Host A and Host B were used, allowing the maximum throughput to reach up to 400 Mbps. Test results using Topology 2, shown in Figure~\ref{fig:plot_isa2}, the measured throughput reached the maximum aggregate bandwidth of 200 Mbps as configured.

\subsection{Evaluation in physical wide-area SDN}
Two virtual machines were used for the evaluation in wide-area SDN. One was deployed in NAIST (Nara Institute of Science and Technology), Japan. Another was deployed in UF (University of Florida). Two paths were used in this experiment. For the first path, NAIST and UF are connected through two physical OpenFlow switches provided by NICT (National Institute of Information and Communications Technology), Japan. For the second path, a GRE link was directly established over the Internet between NAIST and UF.

smoc outperformed POX S-T from the start, then continued to increase its throughput the test as shown in Figure~\ref{fig:plot_ent}. It is noteworthy that since TCP increases window size slowly in wide-area networks due to long round-trip time, more experiment time is needed for smoc to reach the maximum bandwidth possible in the network. Specific to the test in PRAGMA-ENT, we used 12 \texttt{iperf} threads (-P 12) because a larger number of threads would saturate wide-area networks more fully. This means \texttt{iperf} produces more consistent values toward the maximum available bandwidth.

\clearpage
\section{Discussion}
In this section, we discuss the performance of our algorithm,  issues with path installation delay in our controller,  conditions of the test environment, and scalability of our solution.

\subsection{Algorithm performance}
We used a purely topological routing algorithm and generated path sets based on ``minimum shared edges -- minimum hops'' basis. Whlie this is very simple to implement, only requiring a few calculations and no monitoring at all, the performance in real-world WANs may be debatable as the topology alone is not enough to effectively route flows through the best paths. One quick improvement that could be done to this controller is to use bandwidth-based routing by implementing a weighted graph and bandwidth monitoring to supply the graph with weights. Passive bandwidth monitoring was considered because we do not require the level of precision that could only be achieved by active monitoring.
Any changes to the topology in real-time would be noticed by the management modules provided by POX and Overseer.

\subsection{Path installation delay}
We experienced 2-3 second delay in path installation as seen in Figure~\ref{fig:plot_isa}. This delay is caused by the path installation process by underlying POX modules. While this delay may be insignificant when a flow is long enough, it may impact short flows and cause scalability problems when handling a large number of flows. We need to find some way to improve the performance of the controller, such as shifting from the current reactive approach to a more proactive one which is more scalable~\cite{fernandez2013comparing} and has better performance. Some examples of proactive measures possible for smoc include anticipating and preinstalling secondary paths for additional subflows right after the first subflow is created, or storing a group of frequently-used path sets so they do not have to be calculated every time a new flow enters the network.

\subsection{Test environment}
While PRAGMA-ENT is a very good representation of WANs, the segment that we used consist of only two paths and a small number of switches. Even if these switches represented many more actual network elements, a more complex network could prove beneficial to the evaluation of our work. Additionally, testing with real-world applications would provide a realistic picture of our experiment. The high latency present in PRAGMA-ENT caused TCP flows to increase their window sizes more slowly. Shown in Figure~\ref{fig:plot_ent}, it takes about 160 seconds for smoc's TCP flows to collectively increase their throughput to about 160 Mbps. This means spending more time with the test runs on high-latency networks should provide clearer results.

\subsection{Scalability}
Even though the multipath routing algorithm described in this work is adequate to efficiently route a set of subflows belonging to an MPTCP session through multiple paths, the smoc controller itself may have scalability problems. smoc is inherently centralized due to its use of OpenFlow. It has been studied that the number of flows that can be processed by the OpenFlow controller reduces at a quadratic rate with increasing number of switches, regardless of using proactive or reactive approach in routing~\cite{fernandez2013comparing}. As described earlier, reducing path installation delay by using proactive routing and reducing path set computation time can be some simple ways to mitigate (but not completely eliminate) the scalability problem by increasing the rate of flow processing. More involved methods include considering additional features in later OpenFlow versions, such as TCP flag matching introduced in version 1.5.0, to allow the switches to make more decisions on their own without invoking the controller. However, not all switches support the newest versions. We must consider the compatibility between the controller and the target environment carefully before upgrading the protocol version used in our controller.

Apart from the methods mentioned above, we could consider alternatives and modifications to OpenFlow, such as HyperFlow~\cite{hyperflow} and DevoFlow~\cite{devoflow}. HyperFlow uses multiple synchronized OpenFlow controllers to communicate with each other and split the workload. Installing one HyperFlow controller per site may be more scalable than using a single OpenFlow controller for the entire network. On the other hand, DevoFlow, which is a significant modification to OpenFlow, aims to reduce workload on the controller by allowing additional actions on the switches, such as rule cloning and multipath support. These solutions would be able to improve scalability of many existing OpenFlow applications including smoc.

\section{Conclusion}
In this work we presented a simple multipath OpenFlow controller that routes MPTCP sessions by splitting them across multiple paths. Tests on both LAN and WAN SDN testbeds yielded positive results, indicating that our controller works as intended. No modifications to applications or host machines were made (only the kernel in the virtual machines), making our solution backwards-compatible with existing systems. We would find ways to improve its performance in future iterations of our work.

\section{Acknowledgment}
This work was partly supported by JSPS KAKENHI Grant Number 15K00170.
The authors appreciate the collaboration and assistance from the Pacific Rim Applications and Grid Middleware Assembly (PRAGMA) and its members that made the PRAGMA-ENT experimental network testbed which is used in our research possible.
The first author also expresses his gratitude to the Japan Student Service Organization (JASSO), the Japan Ministry of Education, Culture, Sports, Science and Technology (MEXT), and KDDI Foundation for financial support and scholarship during fiscal years 2013, 2014, and 2015.

%
\bibliographystyle{abbrv}
\bibliography{rfc,mptcp,openflow,etc}


%
%
\balancecolumns
\end{document}